\input harvmac.tex

\lref\wya{
Awata, H.,
Kubo, H.,
Odake, S. and Shiraishi, J.:
Quantum\ $W_N$\ algebras
and Macdonald Polynomials.
Commun. Math. Phys. {\bf 179}, 401-416 (1996)} 

\lref\fatt{Fateev, V.A.: The exact relations between the
coupling constants and the masses of particles for the
integrable perturbed conformal field theories.
Phys. Lett. {\bf B324}, 45-51 (1994)}

\lref\weisgt{Coleman, S. and Thun, H.J.: On the
prosaic origin of the double poles in the
sine-Gordon S-matrix.
Commun. Math. Phys. {\bf 61}, 31-39 (1978) }

\lref\mussor{Delfino, G., Simonetti, P. and Cardy, J.L.:
Asymptotic factorization of form-factors in
two-dimensional quantum field theory.
Phys. Lett. {\bf B387}, 327-333  (1996)}

\lref\patrif{Barden, H.W., Corrigan, E., Dorey, P.E.
and Sasaki, R.: Affine Toda field 
theory and exact S matrices. 
Nucl. Phys. {\bf B338}, 689-746 (1990)}

\lref\LZ{Lukyanov, S. and Zamolodchikov, A.:
Exact expectation values of local fields
in quantum sine-Gordon model.
Preprint CLNS 96/1444, RU-96-107,
\#hep-th  9611238}

\lref\FLZZ{Fateev, V., Lukyanov, S.,
Zamolodchikov, A. and
Zamolodchikov, Al.: Expectation values of boundary fields
in the boundary sine-Gordon model.
Preprint CLNS 97/1465, RU-97-04,
\#hep-th  9702190}

\lref\lykyy{Lukyanov, S.: in preparation}

\lref\ff{
Feigin, B. and Frenkel, E.: Quantum\ $ W$-algebras 
and elliptic algebras.
Commun. Math. Phys. {\bf 178}, 653-678 (1996)}

\lref\MIch{Mikhailov, A.V., Olshanetskii, M.A. and
Perelomov, A.M.: Two-dimensional generalized Toda lattice.
Commun. Math. Phys. {\bf 79},  473-488 (1981)}

\lref\Mussar{Mussardo, G.:
Off-critical statistical models factorized scattering
theories and bootstrap program.
Phys. Rep. {\bf 218}, 215-379 (1992)}

\lref\lik{Lukyanov, S.: Form-factors of exponential fields
in the sine-Gordon model. Preprint CLNS 97/1471,
\#hep-th  9703190}

\lref\xorosh{Khoroshkin, S., Lebedev, D. and Pakuliak, S.:
Elliptic algebra\ $A_{q,p}(\widehat{sl_2})$\ 
in the scaling limit. Preprint ITEP-TH-51/96 (1996),
\#q-alg 9702002}

\lref\KouM{Koubek, A. and Mussardo, G.:
On the operator content of the sinh-Gordon model.
Phys. Lett. {\bf B311}, 193-201 (1993)}

\lref\ZaZa{Zamolodchikov, A.B. and Zamolodchikov, Al.B.:
Factorized S-matrices in two dimensions as the exact
solutions of certain relativistic 
quantum field theory models.
Ann. Phys. (N.Y.) {\bf 120}, 253-291 (1979) }

\lref\Luk{Lukyanov, S.: Free field 
representation for massive
integrable models.
Commun. Math. Phys. {\bf 167},  183-226 (1995)}

\lref\Fedya{Smirnov, F.A.: Form-factors in completely
integrable models of
quantum field theory. Singapore: World Scientific 1992}

\lref\LUkyan{Lukyanov, S.: A note on the deformed Virasoro
algebra. Phys. Lett. {\bf 367}, 121-125 (1996)}

\lref\fre{Frenkel, E. and Reshetikhin, N.:
Quantum affine algebras
and deformations of the Virasoro
and\ $W$-algebras. Commun. Math.
Phys. {\bf 178}, 237-266 (1996)}

\lref\yap{
Shiraishi, J., Kubo, H., Awata, H.
and Odake, S.:
A quantum deformation of the
Virasoro algebra and the
Macdonald symmetric functions. Lett. Math. Phys. {\bf 38},
33-51 (1996)}

\lref\lp{Lukyanov, S.  and   Pugai, Ya.: Bosonization
of ZF algebras: Direction toward deformed Virasoro
algebra. JETP {\bf 82}, 1021-1045 (1996)}

\lref\Baxter{Baxter, R.J.: Exactly solved models
in statistical mechanics. London: Academic Press 1982}

\lref\deV{ Destri, C. and  de Vega, H.:
New exact result in affine Toda
field theories: Free energy and
wave-function renormalization.
Nucl. Phys. {\bf B358}, 251-294 (1991)}

\lref\Arien{Arinshtein, A.E., Fateev, V.A. 
and Zamolodchikov, A.B.:
Quantum S matrix of the\ $(1+1)$-dimensional Todd chain.
Phys. Lett. {\bf B87}, 389-392 (1979)}

\lref\FATEEV{Fateev, V.A.: to be published }

\lref\jde{Jimbo, J. and Miwa, T.:
Algebraic analysis of Solvable Lattice
Models. Kyoto Univ., RIMS-981 (1994)}

\lref\yapjj{Jimbo, M., Konno, H. and Miwa T.: Massless
$XXZ$ model and degeneration of the elliptic algebra
${\cal A}_{q,p}(\widehat{sl_2})$. 
Preprint (1996), \#hep-th 9610079 }

\lref\Zamolod{Zamolodchikov, A.B.: unpublished}

\lref\kede{Foda, O., Iohara, K., Jimbo, M., Kedem, R.,
Miwa, T. and Yan, H.: An elliptic 
quantum algebra\ ${\widehat{ sl_2}}$.
Lett. Math. Phys. {\bf 32}, 259-268 (1994)} 

\lref\Clave{Clavelli, L. and Shapiro, J.A.: 
Pomeron factorization
in general dual model. Nucl. Phys. {\bf B57}, 490-535 (1973)}

\lref\daves{Davies, B., Foda. O., Jimbo, M., Miwa, T. and
Nakayashiki, A.: Diagonalization of the $XXZ$ Hamiltonian
by vertex operators. Commun. 
Math. Phys. {\bf 151}, 89-153 (1993)}

\lref\lashkev{Lashkevich, M.: Scaling limit of
the six vertex model in the
framework of free field representation. 
Preprint LANDAU-97-TMP-2, \#hep-th 9704148 } 

\lref\dri{Drinfel'd, V.G.: Quantum groups. In:
Proceedings of the international congress of
mathematics. Berkeley 1986, {\bf 1}, California:
Acad. Press. 1987, 798-820}

\lref\jimb{Jimbo, M.: A $q$-difference analogue of
\ $U( g)$\ and the Yang-Baxter equation. 
Lett. Math. Phys. {\bf 10}, 63-69 (1985)}    

\Title{\vbox{\baselineskip12pt\hbox{CLNS 97/1478}
\hbox{hep-th/9704213}}}
{\vbox{\centerline{
Form-factors of exponential fields}
\vskip6pt
\centerline{in the affine\ $A^{(1)}_{N-1}$\ Toda  model}}}

\centerline{Sergei Lukyanov}
\centerline{Newman Laboratory, Cornell University}
\centerline{ Ithaca, NY 14853-5001, USA}
\centerline{and}
\centerline{L.D. Landau Institute for Theoretical Physics,}
\centerline{Chernogolovka, 142432, RUSSIA}
\centerline{}
\centerline{}
\centerline{}

\centerline{\bf Abstract}
A free field  representation  for
form-factors of  exponential 
operators\ $ e^{{\vec a}{\vec \varphi}}$\ in
the affine\ $A^{(1)}_{N-1}$\ Toda  model is proposed. 
The one  and two particle form-factors 
are calculated explicitly.
\Date{April, 97}

An effective tool for form-factor  calculations
is  so called
"free field representation" approach\ \Luk. 
Its essentials\ \Zamolod\
were borrowed from Baxter's corner
transfer matrix method\ \Baxter,\ \daves,\ \jde. 
The approach is based on
the examination  of dynamical symmetry algebras  
acting  in an angular
quantization space (AQS) of a   massive integrable model.
AQS is a
field-theoretic analog of the space, where  the lattice
corner
transfer matrix acts, and  inherits 
remarkable features  of the later one.
AQS
can be typically identified 
with a representation space
of  some "deformed algebra" taken in
a proper "scaling limit". At
the moment   relations between
deformed algebras and quantum field
theory (QFT) integrable models
are mainly   studied on the 
following examples:

$U_q({\widehat {sl_2}})$ at the level
one \dri, \jimb\ $ \longleftrightarrow$\ 
$SU(2)$ invariant Thirring model \daves, \jde, \Luk,  \lashkev;

Elliptic
algebra\ ${\cal A}_{q,p}
(\widehat{ sl_2})$ \kede\ $\longleftrightarrow$ 
sin-Gordon model\ \Luk,\ \yapjj,\ \xorosh; 

Deformed Virasoro
algebra\ \lp,\ \fre, \yap,\ \ff\ $\longleftrightarrow$ 
sin(h)-Gordon model\ \LUkyan,\ \yapjj.

\noindent
In the last example the dynamical symmetry algebra
manifests itself in  a  very simple way;
A current of the deformed Virasoro algebra
coincides with  the Zamolodchikov-Faddeev operator of  the
basic scalar particle,
acting in AQS\ \LUkyan.
In spite of an  apparent triviality    
of this observation,
it can form the basis
for derivation of form-factors of
local operators. Indeed, using
the bosonization of the 
deformed Virasoro  
algebra\ \yap,\ \ff\ we  can represent form-factors
as traces over   free
Fock spaces\ \Luk.
Then, the   well known technique\ \Clave\ makes
possible to reduce 
the calculation of the traces
to a  vacuum averaging of products  
of vertex operators.
This program have been implemented
in the work\ \lik\ and the
form-factors of 
exponential fields in the sinh-Gordon model\ \KouM\ were
recovered.
The final rules for the  free field representation
appear to be very simple  and 
one can try to generalize 
them to  more complicated QFT models. 
The best candidate to do this
is related  to the 
deformed W-algebra introduced in the works\ \ff,\ \wya.
It would appear reasonable that 
the bosonization
of this algebra provide  
the free field representation for the
form-factors
in the affine Toda field theory.
An analysis of  the results of Refs.\ \ff,\ \wya\ 
shows that currents of the deformed W-algebra in
a proper scaling limit 
coincide with the
Zamolodchikov-Faddeev operators
acting in AQS
of the affine\ $A^{(1)}_{N-1}$\ Toda QFT.
Therefore, the  form-factors of  the model 
can be given
by  traces over AQS and
then rewritten  as   vacuum averages.
The purpose of this letter is to present
the final  rules for the  
calculation of the   form-factors
of exponential fields. 
Using the free field representation, we derive
the one and two particle form-factors and discuss their 
properties.

\newsec{Preliminaries}

The affine\ $A^{(1)}_{N-1}$\ Toda model\ \MIch,\ \Arien\
describes  dynamics of the\ $N-1$\ component 
real scalar field
\eqn\sosoiw{{\vec {\varphi}}(x)=\big(\, {\varphi}^{(1)},...,
{\varphi}^{(N-1)}\, \big)\ ,  }
governed by  the Euclidean action
\eqn\hasy{
{\cal A}_{T} =
\int d^2 x\, 
\bigg\{\, {1\over {8\pi}}\, 
\big(\partial_{\nu}{\vec {\varphi}}\big)^2
+\mu\, \Big(\,  \sum_{k=1}^{N-1}\ 
e^{b{\vec \alpha_k}{\vec\varphi}}+
e^{b{\vec \alpha_0}{\vec\varphi}}\, 
\Big)\, \bigg\}\ .}
Here we denote by 
$${\vec\alpha}_k\in {\bf R}^{N-1}\, ,\ \ \ 
\ {\vec\alpha}_k^2=2\ \ \ \  
(k=1,...,N-1)$$
a set   of the  simple positive roots and
$$-{\vec \alpha}_0=\sum_{k=1}^{N-1}{\vec \alpha}_k $$
is the maximal  root  
of  the simple  Lie algebra $A_{N-1}$. 
The action\ \hasy\ is invariant under the finite
group generated by the elements,
\eqn\juyt{\eqalign{
&{\hat {\cal C}}:\ {\vec \alpha}_k{\vec\varphi}\to
{\vec \alpha}_{N-k}{\vec\varphi}\, ,\  \ \ 
\ \ \ \ \ \ \ {\vec \alpha}_0{\vec\varphi}
\to{\vec \alpha}_{0}{\vec\varphi}\  ;\cr
&{\hat \Omega}:
\ {\vec \alpha}_1{\vec\varphi}\to {\vec \alpha_{2}}
{\vec\varphi}\to
...\to{\vec \alpha}_{N-1}{\vec\varphi}\to
{\vec \alpha_{0}}{\vec\varphi}\to
{\vec \alpha}_1{\vec\varphi}\ .}}
There are $N-1$\ 
particles in the spectrum of the model  and we will denote them   
by $\big\{B_n\big\}_{n=1}^{N-1}$.
Acting on   asymptotic states,\ \juyt\ corresponds to the 
transformations
\eqn\hdydtr{\eqalign{&{\hat {\cal C}}:\ B_n\to B_{N-n}\ ;\cr
&{\hat \Omega}:\ B_n\to \omega^{n}  \,   
B_n\ ,}}
with
$\omega=e^{i{2 \pi \over N} }\ . $
The masses of the particles are simple
related\ \MIch,\ \Arien
\eqn\jsghytqgf{M_n=M\ [ n]\ .}
Here and bellow we use the conventional notation
\eqn\ksisy{[n]={\sin({\pi n\over N})\over
\sin({\pi \over N})}\ .}
The  model\ \hasy\ is an 
integrable QFT. Its two-body\ $S$-matrix,
describing\ $B_n B_m\to B_m B_n$
scattering, was
proposed in\ \Arien. If
\eqn\lsosiu{(p^{\nu}_n+p^{\nu}_m)^2=
M_n^2+M_m^2 +2 M_n M_m \cosh(\theta)\ ,}
is the usual relativistic 
invariant\ \ZaZa, the\ $S$-matrix 
reads
\eqn\fsdsew{{\cal S}_{n m }(\theta)=\prod_{p=1}^{n}
\prod_{q=1}^{m} S\big(
\theta+{i\pi\over N}(n-2  p)-{i\pi\over N}(m-2  q)\big)\  ,}
where
\eqn\ksustr{S(\theta)=
{\sinh({\theta\over 2}+ {i\pi \over N  })\, 
\sinh({\theta\over 2}- {i\pi b\over N Q })\,
\sinh({\theta\over 2}- {i\pi \over N   Q b})\over
\sinh({\theta\over 2}- {i\pi \over N  })\, 
\sinh({\theta\over 2}+ {i\pi b\over N Q })\,
\sinh({\theta\over 2}+ {i\pi \over N   Q b})} }
and
\eqn\swowjiu{Q=b^{-1}+b\ .}
To analyze an  analytical structure of
the two-body S-matrix, it
is useful to rewrite\ \fsdsew\ in the form\ \patrif
\eqn\lsosuy{{\cal S}_{n m }(\theta)
=\prod_{q=1}^{{\rm min}(m,n)}
{F\big(\theta-{i\pi\over N}
-{i\pi\over N}(n+m-2 q)\big)\over
F\big(\theta+{i\pi\over N}+{i\pi\over N}(n+m-2 q)\big)}\ ,}
where
\eqn\sgdfre{F\big(\theta\pm i {\pi\over N}\big)=
{\sinh({\theta\over 2}\pm {i\pi b\over N Q })
\sinh({\theta\over 2}\pm {i\pi \over N   Q b})\over
\sinh({\theta\over 2})
\sinh({\theta\over 2}\pm {i\pi \over  N})}\ .}
In the physical strip\ $0<\Im m\, \theta<\pi$\ the
amplitude\ ${\cal S}_{n m }(\theta)$\ possesses two
simple poles 
at\ $\theta=i{\pi\over N}\, (N-|N-n-m|)\ {\rm and}\ 
\theta=i{\pi\over N}\, |n-m|$.
The first pole
corresponds to  the  particle\ $B_{n+m}$\ (here\ $ n+m $\ 
should be taken modulo $N$) in the direct channel 
$B_n B_m\to B_m B_n$
scattering while the second pole
corresponds to the particle\ $B_{|n-m|}$\ in the cross-channel.
The\ $S$-matrix\ \fsdsew,\ \lsosuy\ has also 
double poles located at
\eqn\uddytr{\theta=i{\pi\over N}\,  \big(\, 
 N- |N-n-m+2 q|\, \big)\, \ \ \ 
q=1,2,...,{\rm min}(m,n)-1\  .}
Simple origin of such singularities has long
been known\ \weisgt.

The  goal of this letter is to propose the free field
representation for the  form-factors
\foot{Our convention for the 
normalization of the asymptotic states
is
$\langle B^{m}(\theta)\,   |\, B_n(\theta')\,\rangle=2\pi
\ \delta_{n}^{m}\,  \delta(\theta-\theta')$.}
$$\langle\, 0\,  |\,  e^{{\vec a}{\vec \varphi}}\, |\, 
B_{n_p}(\theta_p)...B_{n_1}(\theta_1)\, \rangle $$
of  exponential fields\ $e^{{\vec a}{\vec \varphi}}$\ 
in the affine\ $A^{(1)}_{N-1}$\ Toda QFT.
We  assume that the exponential fields are
normalized in accordance with short distance behavior
of the two point correlation function
\eqn\sjsydt{\langle\, 0\,  |\,  e^{{\vec a}{\vec \varphi}(x)}
\, e^{-{\vec a}{\vec \varphi}(x)}\, 
|\, 0\,  \rangle\to |x-y|^{2{\vec a}^2}\, , \ \ \ \ {\rm as}\ \ 
|x-y|\to 0\ .}
Notice that\ $|{\vec a}|$\ should be sufficiently small
number, in order to\ \sjsydt\ be a leading  asymptotic. 
Under the
normalization\ \sjsydt,\ the vacuum  expectation values
\eqn\hstre{{\cal G}_{{\vec a}}=
\langle\, 0\,  |\, 
e^{{\vec a}{\vec \varphi}}\, |\,  0\,  \rangle\ , }
have been recently calculated by V.A. Fateev\ \FATEEV
\foot{For\ $N=2$ the vacuum  expectation values
were found in\ \LZ, \FLZZ.}. 
In this work we do not use their explicit form.
Just note that due to the unbroken discreet
symmetry\ \juyt,\ \hdydtr,
\eqn\sjhdyt{{\cal G}_{{\vec a}}=
1+O({\vec a}^2)\ ,\ \ \ {\rm as}\
\ \ \ \  |{\vec a}|\to 0\ .}

\newsec{Free field representation for form-factors}

The construction given bellow was  
initiated by the results
of Refs.\ \ff\ and\ \wya.

To describe the free field representation 
for the form-factors,
we need to introduce a set of vertex operators
$\big\{\Lambda_{s}(\theta)\big\}_{s=1}^{N}$\
associated with the weights  
${\vec h}_s$\ 
of the first
fundamental (vector) representation\ $\pi_1$\ of $A_{N-1}$
\eqn\sjuyt{\big\{{\vec h}_{s}\big\}_{s=1}^{N}\, :\ \ \ \ \  
{\vec h}_{s}\,  {\vec\alpha}_{k}=\delta_{s, k}-
\delta_{s, k+1}\, ,\ \ \ \ \ \ \ \ k=1,...,N-1\ . }
The averaging  of the products of these
vertex operators will be performed by Wick's theorem using
the rules,
\eqn\jsuyt{\eqalign{
&\langle \langle\, \Lambda_{s}(\theta)
\, \rangle \rangle=1\ ,\cr
&\langle \langle\, \Lambda_{s}(\theta_2) \Lambda_{s}(\theta_1)
\, \rangle \rangle =R(\theta_1-\theta_2)\ ,\cr
&\langle \langle\, \Lambda_{s}(\theta_2)
\Lambda_{s'}(\theta_1)
\, \rangle \rangle =R(\theta_1-\theta_2)\ 
F\big(\theta_1-\theta_2 + {\rm sgn}(s'-s) {i\pi\over N}\big)\, ,
\ \ \ \ \ s\not=s' .}}
Here the function\ $F(\theta)$\ is given
by\ \sgdfre\ and\ $R(\theta)$\ for
$-2\pi<\Im m\, \theta<0$\ reads
\eqn\slffojks{R(\theta)= \exp\biggl\{- 4\int_{0}^{\infty}
{dt\over t} \ {\sinh\big(t(1-{1\over N}\big)\,
\sinh\big({t b\over N Q}\big) \, 
\sinh\big({t \over N  Q b} \big)\over \sinh^2( t)}\ 
\cosh\big( t(1-{i\theta\over\pi})\big)\biggr\}\ }
and  is defined through an 
analytic continuation outside this domain.
$R(\theta)$\ satisfies the relations,
\eqn\ksoooi{\eqalign{&R(-\theta)=S(\theta)\, R(\theta)\, ,   
\ \ \ \ R(-2\pi i-\theta)=R(\theta)\ ,}}
with\ $S(\theta)$\ \ksustr.
{}From the formulas\ \jsuyt\ and\ \ksoooi\ it
follows immediately
that\ $\Lambda_s(\theta)$\ obey the commutation
relations
\eqn\ksidy{\Lambda_s(\theta_1) \Lambda_{s'}(\theta_2)=
S(\theta_1-\theta_2)\ 
\Lambda_s(\theta_1) \Lambda_{s'}(\theta_2)\ .}
They also satisfy the condition
\eqn\ksisuyaa{
:\Lambda_{1}(\theta +
{i\pi (N-1)\over N})\,
\Lambda_{2}(\theta+{i\pi (N-3)\over N})\, ...\,
\Lambda_{N}(\theta-{i\pi (N-1)\over N}):\  =1\ ,}
which is compatible with\ \jsuyt\ due to 
simple properties of the functions\ $F(\theta)\
{\rm and }\  R(\theta)$. 
The normal order dots in\ \ksisuyaa\ $:...:$\ mean
that  we do not 
need to  couple   vertex operators inside the normal
ordered  group under  Wick's averaging.
Weights of the\ $n$-fundamental
representation\ $\pi_n$\  of\
$A_{N-1}$\ are given by the vectors
\eqn\jsytrr{ {\vec h}_{s_1}+...+{\vec h}_{s_n}\,
,\ \ \ \ s_1<s_2<...<s_n\ . }
We associate the vertex operator
\eqn\ksishgfr{ :\Lambda_{s_1}(\theta +
{i\pi (n-1)\over N})\,
\Lambda_{s_2}(\theta+{i\pi (n-3)\over N})\, ...\,
\Lambda_{s_n}(\theta-{i\pi (n-1)\over N}):\  }
with each of these weights.

Let ${\vec \rho }$\ is the  half sum of 
the positive roots of the
Lie algebra $A_{N-1}$,
\eqn\hytr{{\vec \rho }\, 
{\vec \alpha}_k=1\, , \ \ \ k=1,...,N-1\ .}
Introduce the operators
\eqn\hdydre{\eqalign{
{\bf B}_{n}(\theta)=\, &Q\sqrt{{N Z_n\, \over 2 \pi }}\ 
\sum_{1\leq s_1<...<s_n\leq N}\ 
\omega^{({\vec \rho}-{{\vec a}\over Q})
({\vec h}_{s_1}+...+{\vec h}_{s_n})}\ 
\times\cr
&:\Lambda_{s_1}\big(\theta +
{i\pi (n-1)\over N}\big)\, 
\Lambda_{s_2}\big(\theta+{i\pi (n-3)\over N}\big)\,  ...\, 
\Lambda_{s_n}\big(\theta-{i\pi (n-1)\over N}\big):\  .}}
The constants\ $Z_n$\ here will be fixed later
(see (4.16)).
For real\ $\theta$\ these operators 
satisfy the commutation relation,
\eqn\hstr{{\bf B}_{n}(\theta_1) {\bf B}_{m}(\theta_2)=
{\cal S}_{nm}(\theta_1-\theta_2)\ {\bf B}_{m}(\theta_2)
{\bf B}_{n}(\theta_1)\ ,}
where\ ${\cal S}_{n m }(\theta)$\ is given by\ \fsdsew.
We propose the following free field representation for
the form-factors of the  exponential fields
in the affine  $A^{(1)}_{N-1}$ 
Toda QFT,
\eqn\sjsu{
\langle\, 0\, |\,  e^{{\vec a}{\vec \varphi}}\,  |\,
B_{n_p}(\theta_p)...B_{n_1}(\theta_1)\, \rangle=
{\cal G}_{{\vec a}}\ 
\langle \langle\, 
{\bf B}_{n_p}(\theta_p)...{\bf B}_{n_1}(\theta_1)\, 
\rangle \rangle\ .}
This formula reduces the calculation of
the  form-factors
to a combinatoric procedure.
For\ $N=2$\ \sjsu\ turns out to be 
the free field representation
proposed in the work\ \lik\ and reproduces the
results of\ \KouM.

\newsec{One particle form-factors}

{}From the formula\ \sjsu\ we immediately    have
the expression for the  one particle form-factors,
\eqn\ksjhsu{
\langle\, 0\, |\,  e^{{\vec a}{\vec \varphi}}\,  |\,
B_{n}\, \rangle={\cal G}_{{\vec a}}\ 
Q\sqrt{{N Z_n\, \over 2 \pi }}
\ {\chi_n}\big(
{ {\vec a}\over Q}\big)\ .}
The  function\ ${\chi_n}(
{\vec \lambda})$\ coincides with the character of
the $n$-fundamental representation $\pi_n$,
\eqn\jdy{\chi_n(
{\vec \lambda})={\rm tr}_{\pi_n}
\Big[\,  \omega^{({\vec\rho}-{\vec\lambda}) {\vec H}}\,
\Big]\ .} 
Here\ $ {\vec H}=(H^{(1)},...,H^{(N-1)})$\ is
a basis in the  Cartan subalgebra of $A_{N-1}$,
normalized with
respect the Killing form\ $\langle\, ,\, \rangle: \, \langle\,
H^{(\sigma)} ,H^{(\sigma')}\, \rangle=\delta^{\sigma \sigma'}$.
One can show that
\eqn\hsgdre{{\chi_n}(
{\lambda{\vec \alpha_j}})=  4\,
\sin\big(  {\pi\lambda\over N}\big)\,
\sin\big(  {\pi\over N}(\lambda-1)\big)\,
\omega^{-j n}\ [n]\,  \ \ \ \ j=0,...,N-1\ .}
Eqs. \hsgdre,\ \sjhdyt\ allow one 
to specify  the one particle
form-factors of the field\ ${\vec \varphi}$\ by taking the
small\ $|{\vec a|}$\ limit of\ \ksjhsu,
\eqn\sksiy{\langle\, 0\, |\, {\vec \alpha}_j
{\vec   \varphi}\,  |\, B_n\,
\rangle=-2 \sqrt{{2\pi\over N}}\ \omega^{-j n}\,
\sin\big({\pi n\over N}\big)\,  \sqrt{Z_n}\, ,
\ \ \ \  j=0,...,N-1\  .}
In writing\ 
\sksiy\ we assume that the field\ \sosoiw\ is  normalized
in accordance with the short distance limiting behavior,
\eqn\sywtr{\langle\, 0\, |\,
\varphi^{(\sigma)}(x) \varphi^{(\sigma')}(y)\, |\, 0\, 
\rangle\to
-2 \delta^{\sigma \sigma'}\ \log( M |x-y|  )+ O(1)\ ,
\ \ {\rm as}\ \
|x-y|\to 0\ .}
Hence\ $ Z_n$\ in the
definition\ \hdydre\ are the wave-function
renormalization constants  
in the affine\ $A^{(1)}_{N-1}$\ Toda QFT.

With \ \ksjhsu\ and\ \hsgdre\ we can check that
the one particle  form-factors
agree with the quantum equations of motion
\eqn\hsystr{\partial_{\nu} \partial^{\nu} \big(
\,{\vec\alpha_j}{\vec\varphi}\,\big)={\cal M}\, \Big(
2 e^{ b{\vec \alpha}_j{\vec\varphi} }-
e^{ b{\vec \alpha}_{j-1}{\vec\varphi}}-
e^{ b{\vec \alpha}_{j+1}{\vec\varphi}}\Big)\ ,}
with
$${\cal M}={\pi M^2\over 4  Q N {\cal G}\,  \sin\big(
{\pi \over N bQ}\big)
\sin\big({\pi b\over NQ}\big) 
\sin\big({\pi\over N}\big)}\ ,$$
and \ ${\cal G}\equiv{\cal G}_{b{\vec\alpha_j}}\, 
(j=0,...,N-1)$.

Eq.\ \hsystr\ can be used  in derivation of
the specific free energy\ $f(\mu)=-{1\over V}\,
\log Z$\ in the affine\ $A^{(1)}_{N-1}$\ Toda QFT.
According to\ \hasy, the
constant\ ${\cal M}$\ should coincide with\ $4\pi\mu$. 
If we combine this requirement with the obvious relation
\eqn\kssiu{N \, {\cal G}=-\partial_{\mu} f(\mu)\ ,}
we get the formula
\eqn\ksisyr{b Q\ \mu\, 
\partial_{\mu}f(\mu)=-{M^2(\mu)\over 16
\sin\big(
{\pi \over N bQ}\big)
\sin\big({\pi b\over NQ}\big) 
\sin\big({\pi\over N}\big)}\ .} 
Due to \sjsydt,\ $M^2(\mu)\sim \mu^{{1\over b Q}}$\ up 
to dimensionless
constant
\foot{This constant was calculated in the work\ \fatt.}
and we reproduce the  
result of C. Destri and H. de Vega\ \deV
\eqn\ksiyt{f(\mu)=-{M^2(\mu)\over 16
\sin\big(
{\pi \over N bQ}\big)
\sin\big({\pi b\over NQ}\big) 
\sin\big({\pi\over N}\big)}\ .}

\newsec{Two particle form-factors}

In order to simplify the notation in this section
we assume that integers\ $m\leq n, \ m+n\leq N$\ without 
loss in generality. 

By applying the free field representation\ \sjsu, the 
two particle form-factors were derived.
The computations are  essentially   based on the
interesting trigonometric
identity
\eqn\uylki{\sum_{1\leq k_1 \leq...\leq k_m\leq n+1}\ 
\prod_{p=1}^m\, F\big(\theta+
{i\pi\over N} (n-m-2 k_p+2 p+1)\big)=
\sum_{k=0}^m \ C_k^{m+n}\ {\cal K}_{m-k}^{n-k}(\theta)\ ,}
where\ $C_k^{m+n}$\ are the binomial coefficients 
and the functions\  ${\cal K}_{m}^{n}(\theta)$\ are 
defined by
the  following
recursion 
\eqn\jsustre{\eqalign{&\ \ \ \ \ \ \ \ \ \
\ \ \ \ \ \ \  
{\cal K}_{0}^{n}(\theta)\equiv 1\ ,\cr
{\cal K}_{m}^{n}(\theta)=-&
{\sin\big({\pi b\over N Q}\big)
\sin\big({\pi \over N Q b}\big)\ [m+n]
\over \sinh\big({\theta\over 2}+i{\pi(m+ n)\over 2 N}\big)
\sinh\big({\theta\over 2}-i{\pi(m+ n)\over 2  N}\big)}\
\sum_{k=1}^{m}  
{\cal K}_{m-k}^{n-k}(\theta)\ {[ k]\,  [m+n-k]\over
[m]\, [n]}\ .}}
Eq. \uylki\ along with simple properties of symmetric 
polynomials
\foot{The characters\ \jdy\ are symmetric polynomials
with respect of\ $\big\{\omega^{({\vec \rho}-
{{\vec a}\over Q}){\vec h}_s}\big\}_{s=1}^{N}\ $.},
makes it possible to derive the formula\
$(m\leq n, \ m+n\leq N)$
\eqn\kiuytr{\eqalign{&
\langle\, 0\, |\,  e^{{\vec a}{\vec \varphi}}\,  |\,
B_{m}(\theta_2)B_{n}(\theta_1)  \, \rangle=
{\cal G}_{{\vec a}}\
{Q^2 N\over 2 \pi }\, 
\sqrt{ Z_n Z_m} 
\times\cr
&{\cal  R}_{ nm}(\theta_1-
\theta_2)
\  \sum_{k=0}^{m}\ \chi_k\big({{\vec a}\over Q}\big)
\chi_{n+m-k}\, 
\big({{\vec a}\over Q}\big)\ 
{\cal K}_{m-k}^{n-k}(\theta_1-\theta_2)\ .}}
Here\ $  \chi_{n}({\vec\lambda})$\ are the  characters of
the\ $n$-fundamental 
representations\ \jdy\ $(\chi_{0}({\vec\lambda})\equiv 1$)\ 
and
\eqn\uytaas{{{\cal R}_{nm}}(\theta)=
\prod_{p=1}^{n-1}\prod_{s=1}^{m}
F\big(\theta +i{\pi\over N} (n-m-1-2p+2 s)\big)\
\prod_{p=1}^{n}\prod_{s=1}^{m}
R\big(\theta +i{\pi\over N} (n-m-2p+2 s)\big)\ .}
Notice that the function\ $ {{\cal R}_{nm}}(\theta)$\
is  so called "minimal 
form-factor"\ \deV. It admits  similar 
to\ \slffojks\ representation
\eqn\fojks{\eqalign{
&\ \ \ \ \ \ \ \ \ \ \ \ \ \ \ \ \ \ \ \ 
\ \ \ \ \ \ \ \ \ \ \ \ \ \ {{\cal R}_{ nm}}(\theta)=\cr
& \exp\biggl\{- 4\int_{0}^{\infty}
{dt\over t} \ {\sinh\big({tm\over N}\big)\,
\sinh\big(t(1-{n\over N})\big)\,
\sinh\big({t b\over N Q}\big) \,
\sinh\big({t \over N  Q b} \big)\over \sinh\big({t\over N}\big)
\sinh^2( t)}\
\cosh\big( t(1-{i\theta\over\pi})\big)\biggr\}\ .}}
For\ $m+n>N$\ the form-factors can be obtained from\ \kiuytr\
by the use of
the  charge  conjugation
${\hat {\cal C}}$\ \juyt,\ \hdydtr.
Taking\ $|{\vec a}|\to 0$\ limit of\ \kiuytr,
we derive the two particle form-factors
of the field\ $\vec \varphi$
\eqn\gstsreaaa{\eqalign{
\langle\, 0\, |\,  {\vec \varphi}\,  |\,
B_{m}(\theta_2)B_{n}(\theta_1)  \, \rangle&=
\langle\, 0\, |\,  {\vec \varphi}\,  |\,
B_{m+n}\, \rangle\ \sqrt{ {Z_n Z_m\over Z_{n+m}}}\ 
Q\ \sqrt{{N \over 2 \pi}}\times\cr  
&{\cal R}_{ nm}(\theta_1-
\theta_2)\
{\cal K}_{m}^{n}(\theta_1-\theta_2)\ .}}
A simple check based on\ \jsustre\ and\ \hsgdre\ shows 
that Eqs.\ \kiuytr, \gstsreaaa\ conform
with  the  quantum equations of motion\ \hsystr.

Let us discussed basic  properties of the  two particle
form-factors.
First of all, we consider 
their leading
behavior
in the  large\ $|\theta_1-\theta_2|$\ limit.
Using the formulas
\eqn\gstse{R(\theta)\to 1\, , \ \ \ 
\ \ F(\theta)\to 1\ ,\ \ \ 
{\rm as}\ \ \ \ \theta\to\infty\ ,}
we  obtain the cluster property\ \KouM,\ \mussor
\eqn\rwqiu{\langle\, 0\, |\,  e^{{\vec a}{\vec \varphi}}\,  |\,
B_{m}(\theta_2)B_{n}(\theta_1)  \, \rangle\to\ 
{\langle\, 0\, |\,  e^{{\vec a}{\vec \varphi}}\,  |\,
B_{m} \, \rangle 
\langle\, 0\, |\,  e^{{\vec a}{\vec \varphi}}\,  |\,
B_{n}\, \rangle\over
\langle\, 0\, |\,  e^{{\vec a}{\vec \varphi}}\,  |\,
0\, \rangle}\ ,}
as\ $ |\theta_1-\theta_2|\to \infty$\ while
\eqn\hsystresewq{\eqalign{
\langle\, 0\, |\,  {\vec \varphi}\,  |\,
B_{m}(\theta_2)B_{n}(\theta_1)  \, \rangle&\to
-\langle\, 0\, |\,  {\vec \varphi}\,  |\,
B_{m+n}\, \rangle\ 
\sqrt{ {Z_n Z_m\over Z_{n+m}}} \times\cr
&2\,  Q\sqrt{{2 N \over  \pi}}\ 
\sin\big({\pi b\over N Q}\big)
\sin\big({\pi \over N Q b}\big)\ 
[m+n]\ e^{-|\theta_1-\theta_2|}
\ .}}
The last relation  was 
established in the work\ \deV\ by  an
analysis  of the  
Feynman diagrams.

The form-factors\ \kiuytr, \gstsreaaa\ satisfy 
Watson's and crossing symmetry equations\ \Fedya\ and
have required  analytical structure\ \deV.
Specifically,  they  develop simple 
poles at the
points
$$\theta_2-\theta_1=i {\pi\over N}\, (m+n-2 q)\,
 \ \ \ \ ( q=1,2,...,m-1) \ . $$
The latter   do not correspond to
bound states and are  directly  related to the
double pole structure of the two-body\ $S$-matrix \uddytr.
At the same time, the form-factors  singularity   
at\ $  \theta_2-\theta_1=
i{\pi\over N} (m+n)\, ,$
\eqn\osisusy{
\langle\, 0\, |\,  e^{{\vec a}{\vec \varphi}}\,  |\,
B_{m}(\theta_2)B_{n}(\theta_1)  \, \rangle\to
{i\ \Gamma_{  nm}\over \theta_2-\theta_1-i{\pi\over N} (m+n)}\ \  
\langle\, 0\, |\, e^{{\vec a}{\vec \varphi}}\,  |\,
B_{m+n}\, \rangle\   ,}
is in agreement with the  bound state pole of\ \fsdsew.
To find the  residue it  is useful to take advantage of
the identity\ \uylki,
\eqn\trewsd{\Gamma_{ nm}=   
\sqrt{ {Z_n Z_m\over Z_{n+m}}}\   Q\, \sqrt{{2 N \over  \pi}}\
{\sin\big({\pi b\over N Q}\big)
\sin\big({\pi \over N Q b}\big)
\over \sin\big({\pi \over N} \big)}
\ {\cal R}_{nm}\big(-{i\pi\over N} (m+n)\big)\, 
\prod_{q=1}^{m-1} F\big(i{\pi\over N} (2 q+1)\big)\ .}
According to  the common requirement\ \Fedya,
$\Gamma_{ nm}$\ should also define
the residue of the 
two-body\ $S$-matrix in the following manner,
\eqn\hsystr{{\cal S}_{nm}(\theta)
\to { i\ \big(\Gamma_{ nm}\big)^2\over
\theta-i{\pi\over N} (m+n)}\, ,\ \ \ {\rm as}\ \ \
\theta\to i{\pi\over N} (m+n)\ .}
Using\ \lsosuy,\ we conclude that
\eqn\hsystr{\big(\Gamma_{ nm}\big)^2=
2\, {\sin\big({\pi b\over N Q}\big)
\sin\big({\pi \over N Q b}\big)
\over \sin\big({\pi \over N} \big)}
\ {\prod_{q=1}^{m-1} F\big(i{\pi\over N} (2 q+1)\big)\over
\prod_{q=1}^{m} F\big(i{\pi\over N} (2n+2 q-1)\big)}\ .}
Relations\ \trewsd\ and\ \hsystr\
restrict\ $Z_{n}$\ up to  a $n$-independent constant\ $c$,
\eqn\terww{Z_{n}= {\pi\ c^n\   \sin\big({\pi \over N }\big)
\over N\,  Q^2\ \sin\big({\pi b\over N Q}\big)
\sin\big({\pi \over N Q b}\big)}\ 
\prod_{p=1}^{n-1}\, { \Big[\, R\big(-{2\pi i \over N} p\big)\, 
F\big({i\pi \over N} (2 p+1)\big)\, \Big]^{2n-2p}\over
F\big({i\pi \over N} (2 p+1)\big)}\ .}
If we  enclose also the condition 
\eqn\kdidy{Z_{n}=Z_{N-n}\ ,}
which follows from the
invariance with respect to  the  charge  conjugation\ 
${\hat {\cal C}}$\ \juyt,\ \hdydtr,
the constant\ $c$\ is specified uniquely.  
Finally  we have
\eqn\solsdsi{\eqalign{
Z_n=&{\pi \over Q^2 \sin({\pi b\over Q})}\times\cr
&\exp\biggl\{2\int_{0}^{\infty}
{dt\over t} \ {\sinh\big({t b\over N Q}\big) \,
\sinh\big({t \over N  Q b} \big)\over \sinh^2( t)
\sinh({ t\over N})}\
\Big(\sinh^2\big({t n\over N }\big)+
\sinh^2\big(t (1- {n\over N} ) \big) \Big) \biggr\}\ . }}
The expression\ \solsdsi\ for the  wave-function
renormalization constants
was previously obtained in the
work\ \deV. It matches  all necessary field-theoretic
bounds, including the one loop perturbative check.
In such a manner we conclude 
that the explicit formulas for the  one and two
particle form-factors
support
the proposed free field representation.

In closing  we note that  the  free field representation
admits straightforward generalization for any  untwisted
affine\ $ ADE$\ Toda QFT\ \lykyy.

\hskip1.0cm

\centerline{\bf Acknowledgments}

This work 
is supported in part by NSF grant.

\listrefs

\end